# An Efficient Data Warehouse for Crop Yield Prediction

**Vuong M. Ngo, Nhien-An Le-Khac, M-Tahar Kechadi**

School of Computer Science, College of Science,

University College Dublin, Belfield, Dublin 4, Ireland

*Abstract.*

*Nowadays, precision agriculture combined with modern information and communications technologies, is becoming more common in agricultural activities such as automated irrigation systems, precision planting, variable rate applications of nutrients and pesticides, and agricultural decision support systems. In the latter, crop management data analysis, based on machine learning and data mining, focuses mainly on how to efficiently forecast and improve crop yield. In recent years, raw and semi-processed agricultural data are usually collected using sensors, robots, satellites, weather stations, farm equipment, farmers and agribusinesses while the Internet of Things (IoT) should deliver the promise of wirelessly connecting objects and devices in the agricultural ecosystem. Agricultural data typically captures information about farming entities and operations. Every farming entity encapsulates an individual farming concept, such as field, crop, seed, soil, temperature, humidity, pest, and weed. Agricultural datasets are spatial, temporal, complex, heterogeneous, non-standardized, and very large. In particular, agricultural data is considered as Big Data in terms of volume, variety, velocity and veracity.*

*Designing and developing a data warehouse for precision agriculture is a key foundation for establishing a crop intelligence platform, which will enable resource efficient agronomy decision making and recommendations. Some of the requirements for such an agricultural data warehouse are privacy, security, and real-time access among its stakeholders (e.g., farmers, farm equipment manufacturers, agribusinesses, co-operative societies, customers and possibly Government agencies). However, currently there are very few reports in the literature that focus on the design of efficient data warehouses with the view of enabling Agricultural Big Data analysis and data mining. In this paper, we propose a system architecture and a database schema for designing and implementing a continental level data warehouse. Besides, some major challenges and agriculture dimensions are also reviewed and analysed.*

## 1. Introduction

Annual world cereal production in 2017 is estimated to be 2.64 billion tonnes (FAO – 2017) and by 2050 will need an increase of 60% to meet global population growth, which is forecasted to reach a total of 9.8 billion (Alexandratos and Bruinsma – 2012, United Nations – 2017). Hence, to satisfy the massively increased demand for food, crop yields must be significantly increased by using new farming technologies or methods. In agriculture, the key questions that we want to answer were always there since the beginning, maybe with some different constraints. These questions include (1) what kind of crops are suitable to a given field? (2) How to improve the crop yields for a given season? (3) Where can we sell at the highest price? etc. Until recently, it was not easy to access suitable information in order to answer such questions. That is because, they require the combination of many specific areas of knowledge and information on agriculture, technology, market and customer behaviour. For instance, for the first question, one needs a deep understanding of different crops characteristics and information about soil and weather thus allowing one to associate crops to suitable types of soil, weather conditions and market needs. For the second question, one needs to know the right time to plant it; depending on the period of crop development, water delivery, pesticide spray, and fertiliser calibration. As for the final question, we usually need the crop price, the supply, and demand information within the last few years, etc.

With technological advances in the area of information and communication technology (ICT), farmers can access and share valuable information and knowledge. They can also obtain knowledge on new research results about seeds, reduction in post-harvest losses, finding better ways to access markets and so on. Besides, they can participate intensively in policy dialogues and discussions. More importantly, they can use some online tools to gather prices from different sources and access forecasted price and possible market for crop outputs (World Bank Group 2017).

Risks and uncertainty are common in agriculture most of which are due to irregular weather (such as droughts and floods), market movements and pest outbreaks. Recent advances in ICT can drastically reduce the costs of collecting, processing and disseminating information to supply early warnings to farmers. For instance, sensor technologies are currently used to gather data from the fields to estimate, forecast, or monitor damage. Timely and early warnings are essential to limit large losses. Additionally, ICT can provide efficient instruments for managing and mitigating risks.

In 2016, the production of cereals in the 28-member countries of the European Union (EU-28) was 301 million tons which represented about 11.6% of world-wide cereal production. EU-28 is ranked 3$^{rd}$ in the world after China and United States (Momagri report). There are about 10 million people working in agricultural section in the EU-28, which accounts for 4.4% of the total EU-28 employment (Jortay-2017). In Strandell and Wolff (2016), the total agricultural area was about 186.4 million hectares which occupies 42.5% of the EU-28's total land area. This is higher than the average of the world total agricultural area which is 37.9%. For EU agricultural land, arable land (including cereal land), grassland (including pasture, meadow and rough grazing) and permanent crops (such as fruits, berries, nuts, citrus, olives and vineyards) account for 59.8%, 34.2% and 6%, respectively. The five most important crops in Europe are wheat, grain maize, barley, oats and winter cereal.

As reported in (Eurobarometer report - 2016), it was agreed that agriculture and rural areas were very important for the future, and that agriculture and rural development budgets should be increased. It was stated in the same report that precision agriculture (PA) is of vital importance to the future of agriculture in Europe. They found that PA can make a significant contribution to food security and safety. Besides, PA will develop new ways of farming and will have direct effect on the society (Schrijver - 2016). PA was adopted in 1997 and since then it has shown continual increases in adoption.

In this research, we focus on the development of an agricultural data warehouse for building a precision agricultural system. Building such a data collection and storage system which is suitable for rapid analytics is one of the important steps within the agricultural data analysis process which will can be used subsequently for agronomic decision-making. The benefits here include improvement of crop yield, lower input of chemicals, better farm management, and marketing to meet the world's future needs.

Accordingly, we firstly review major challenges for developing efficient data warehouses for agronomic purposes and then present methods for designing and implementing a continental data warehouse. Some of the challenges include 1) multiple sources of data, which are heterogeneous with varying levels of details and quality, 2) the issue of high dimensionality and its impact on the design choices and storage models, 3) definition of efficient and optimized schemas in terms of both data access and storage space, 4) the issue of interoperability with standards in the field.

The rest of this paper is organised as follows: in the next Section we review the related work in the above context and Section 3 covers the application of ICT to the agricultural domain. We then present our approach and describe the architecture, challenges and schema of the proposed data warehouse in Section 4. Finally, Section 5 gives some concluding remarks.

## 2. Related Work

Data mining being at the heart of data analytics can be used to design an analysis process for exploiting big agricultural datasets. Recently, many papers have been published that exploit machine learning algorithms on sensor data and build models to improve agriculture economics. In Pantazi, et al (2016), the authors used supervised learning models to predict crop yield. Their models are based on self-organizing-maps, namely, supervised Kohonen networks, counter-propagation artificial networks and XY-fusion. They associated high resolution data on soil and crop with iso-frequency classes of wheat yield productivity. In Park, et al. (2016), three rule-based machine learning; namely random forest, boosted regression trees, and Cubist, were used to predict drought conditions. The proposed models identify the relative importance of drought factors which varies by region, time, and drought type. The drought factors were collected from Moderate Resolution Imaging (MRI) Spectroradiometer and Tropical Rainfall Measuring Mission satellite sensors. In addition, Feng et al (2017) proposed two models that used Extreme Learning Machine and Generalized Regression Neural Network. The models estimated water evapotranspiration based on daily temperature data which were obtained from six meteorological stations. Moreover, satellite data were also used to estimate grassland biomass (Ali, et al – 2017). The authors used Multiple Linear Regression, Neural Networks, and Adaptive-Neuro Fuzzy Inference models to estimate a such factor.

There are also papers that have proposed "smart agriculture frameworks". Golubovic et al (2016) proposed a framework to aggregate data inputs and notify when some key events are likely to occur, such as difficult weather conditions and the spread of pests. In their framework, users could provide areas and events of interests by keywords. The systems could return related information published in social media sources like Twitter. Shekhar *et al.* (2017) presented general comments and advice to support smart agriculture infrastructure, which are shared data spaces and intelligent cyber-infrastructure. The authors also presented interdisciplinary research to envision and test the feasibility of the next generation infrastructure. From a Big Data point of view, Schnase *et al.* (2017) used modern infrastructures, such as Cloud Computing, Hadoop, MapReduce, and Cloudera, to build a big data management and analysis platform. The platform provides a high-performance analytics and scalable data management in climate research. Kamilaris et al (2018) proposed software platform which uses Big data analysis on environmental sensor data about land, water and biodiversity. These technologies were Apache Hive and Hadoop and were used for storing and analysing data, respectively. This platform can support farmers on decision-making and administrators in policy-planning, which help increase food production with lower environmental impact.

The above papers did not discuss how to build a data warehouse for a precision agriculture. Two quite old papers – Schulze, *et al.* (2007) and Nilakanta, *et al.* (2008) – presented ways of building a data warehouse for agricultural data. In Schulze, *et al.* (2007), the authors extended Entity-Relationship Model for modelling operational and analytical data which is called the multi-dimensional Entity-Relationship Model. They introduced new representation elements and showed the extension of an analytical schema. Thought, as the schemas were based on entity-relationship model they cannot deal with high-performance, which is the key feature of a data warehouse. In Nilakanta, *et al.* (2008), the authors presented a data warehouse architecture for smart agriculture. They used a star schema model and each fact table has an individual star schema that creates a data mart. All data marts are connected via some common dimension tables. However, the data warehouse concerns livestock farming, not for crop farming. In addition, the number of dimensions in the tables is very small; only 3-dimensions – namely, Species, Location, and Time. Moreover, a star schema is not enough to present complex agricultural information and it is difficult to create new data marts for data analytics.

## 3. Applications of Precision Agriculture

Precision agriculture (PA) or precision farming is a set of techniques created by combining information technology and agriculture science to increase crop yield. Precision agriculture can be defined as the matching of agronomic inputs and practices to localised conditions in field and the improvement of the accuracy of their application (Finch, Samuel and Lane - 2014). The farming management in PA uses information technology, satellite positioning, remote sensing and proximal data gathering as observations measuring and responding to inter and intra-field variability in crops (JRC - 2014).

The first wave of PA's mission was how to better understand soil, fertility, genetic and the use of satellite positioning systems. In fact, the development of sensor and Global Navigation Satellite System (GNSS) technologies improved significantly farming management tasks, such as cultivation, seeding, fertilization, herbicide application, and harvesting. These farm management systems can also assess the status of soils, record weather information, quantify the nutrient status of crops, etc. The newest wave of PA's mission is to combine information technology and agricultural science to build a decision-support system driven by data collected from yield monitors and multifunction tractor displays. PA promises both high quantity and quality of its products with minimum of resource usage, such as water, energy, fertilisers, and pesticides (Schrijver – 2016). These promises will significantly reduce the cost, reduce environmental impact, and produce much better food products.

In the World Bank Group paper (2017), many applications of ICT to Agriculture are described. Land markets and land reform activities were more effectively supported by land information infrastructure. The infrastructure can also integrate land administration services into the wider e-government arena. The forest management systems involve more public participation for better protection and conservation. With the development of Internet of Thing technologies, farms are more connected to collect and analyse data in real-time.

Moreover, Controlled Traffic Farming (CTF) is a PA technology used widely on arable land. CTF avoids unnecessary crop damage and soil compaction. Besides, the method also avoids repetitive agricultural vehicle passes on the field to reduce fuel use and improve timeliness of operations. The environmental benefits of using CTF are many; one can cite reduction in surfaced nutrient leaching and reduction in fertilisers and pesticides quantities. Another PA method which helps optimising the use of fertilisers is Variable Rate Application (VRA). VRA, based on its spatial and temporal components, allows the use of more precise amount of fertilisers in arable lands (JRC-2014).

In fruit farming, automatic systems based on machine vision can monitor food quality and safety, such as colour, size, shape, sugar content and other features. For example, determining the harvest time is very important for grape quality to avoid mixing different grapes' types and maturity. Thus, precision irrigation systems are being used to save water while improve crop yield and quality. These can avoid regulated deficit irrigation, partial root drying and sustained deficit irrigation. In Cohen (2017), the Gavish system of Israel can control the irrigation either separately or together with liquid fertilisers. The system is efficient for saving water and fertilisers while maximising crop yield and cutting production costs.

There are other maturing technologies which are finding application in agriculture. Drones can monitor crop health. Soil monitoring systems can assist in tracking and improving the quality of physical, chemical, and biological properties of soil. Smart greenhouses allow to plant crops with minimal human intervention by automated actions which maintain optimal conditions of temperature, humidity, luminosity and soil moisture (XLabs report-2017).

Eventually, the use of Big Data approaches can provide precise information for more control of farming efficiency and waste. There are six popular areas: (1) Awareness: know people's opinions by using sentiment analysis; (2) Understanding: learn why something is happening, for instance, why food prices have gone down or why water shortages have arisen; (3) Advice: provide targeted advice to specific farms; (4) Early warning: identify early problems, such as disease or pest outbreaks; (5) Forecasting: predict future trends, such as prices for specific crops in specific areas; (6) Financial services: provide credit and insurance to farmers who need them.

## 4. Data Warehouse for Precision Agriculture

### 4.1. Multiple Sources of Data and Challenges

The input data for the PA data warehouse used here were primarily obtained from an agronomy company including data from its operational systems, research results and field trials. These datasets were collected from demonstration farms, tens of thousands of field trials and this data comes from several European countries. There is a total of 29 datasets. On average, each dataset contains 18 tables and is about 1.4 GB in size. Some important tables that are defined for the PA data warehouse are described in Section 4.4.

The main objective of the research is to build a system to efficiently handle agricultural data and facilitate the deployment of practical data mining techniques (Le-Khac et al. 2010) on these datasets with the view to extract useful knowledge and support decision-making. The available datasets are of two types endogenous and exogenous datasets. The endogenous data was collected and provided directly by the company, while the exogenous data concerns the external sources, such as weather systems, farmers, PA consultants, government agencies, retail agronomists and seed companies. They can help with information about local weather, pest and disease outbreak tracking, crop monitoring, market accessing, food security, products, prices and knowledge. Since individuals could be asked to share these information, user privacy and data security become important issues for the system. All these datasets need to be aggregated before mining them.

PA presents many challenges in building an efficient data warehouse because the data is very big and very diverse. This is part of Big Data challenge. The main features of Big Data were first described by Pettey and Goasduff (2011); volume, variety, velocity, and veracity. The amount of agriculture data is rapidly increasing and is intensively produced by different sources, such as trial results, weather data, operational systems, sensor information, satellite images, and so on. Agricultural data has many different forms and formats, such as structured and unstructured data, imagery, metrics, geo-spatial, text, multi-media, model and equation. The velocity feature is the measure of how fast the data is produced and collected, as sensing technologies and other mobile devices are becoming more efficient and cheaper. Therefore, cleaning, aggregating and harmonising datasets in real-time for requirements of analytic operations becomes one of the key challenges. Another difficult challenge is the veracity of agricultural datasets since it is very important to have high quality data before analysing it. Tendency of agronomic data is uncertain because the data is gathered from sensors and manual processes. The heterogeneous sources often show inconsistency, ambiguity, latency and errors in the data. Finally, the PA system can be used by different kinds of users at the same time, for instance by both farmers and agronomists. Every type of user needs to query different information sets thus requiring specific analytics. All of these issues and constraints make the design of large scale data warehouses very challenging.

### 4.2. DW Architecture

In general, a data warehouse (DW) is a federated repository for all the data that an enterprise can collect through different data sources. The data sources can be various enterprise's business systems or external inputs. Golfarelli and Rizzi (2009) and Inmon (2005) defined data warehouse as a collection of data that supports decision making processes. It is defined around key subjects and its data should be integrated, consistent, evolving over time and non-volatile.

A typical data warehousing architecture which is illustrated Figure 1 has four separate and distinct modules: Raw Data (Data Sources), Extraction Transformation Loading (ETL), Integrated Information, and Data Mining (Kimbal and Ross – 2013). The Raw Data module is originally stored in operational databases systems or comes from external information systems. The raw data requires cleansing to ensure data quality before being used. The ETL module includes tools for extracting, transforming and loading the data into a data warehouse storage.

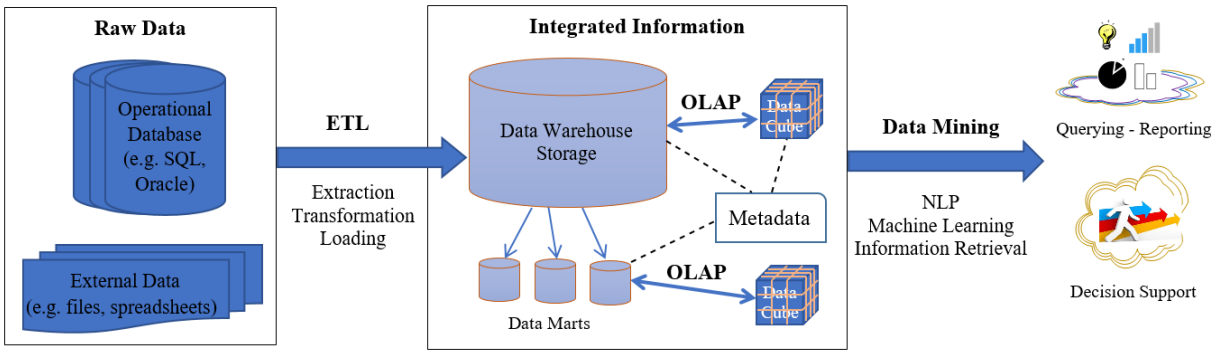

Figure 1. Data warehouse architecture

The Integrated Information module which is a logically centralised repository, contains the data warehouse storage, its metadata, and its OLAP engine. The data warehouse storage is organised, stored and accessed using a suitable schema defined in the metadata. Often the enterprise data warehouse is generated from a set of data marts. A data mart is a subset of the data warehouse storage and is usually oriented to a particular business function or department. The data is usually extracted in a form of data cube before it is analysed in the data mining module. A data cube is a data structure that allows fast analysis of data according to the multiple dimensions that define a business problem. The data cubes are created by online analytical processing engine (OLAP). Details of OLAP and data warehouse schema will be described in Sections 4.3 and 4.4. Finally, Data Mining module contains a set of learning techniques for data analysis and knowledge extraction. The knowledge is represented in a form that it can be interpreted quickly and easily by the users. It is usually represented graphically or as rules.

### 4.3. Quality Criteria

The accuracy of data mining and data analysis techniques depend on the quality of DW. As mentioned in Adelman and Moss (2000) and Kimball and Ross (2013), a quality DW should meet the following important criteria:

- Making information easily accessible.
- Presenting information consistently.
- Integrating data correctly and completely.
- Adapting to change.
- Presenting information in a timely way.
- Being a secure bastion that protects the information assets.
- Serving as the authoritative and trustworthy foundation for improved decision making. The analytical tools need to provide right information at the right time.
- Achieving benefits, both tangible and intangible.
- Being accepted by users using the DW.

The above criteria must be formulated to become measurements. For example, with the last criterion, a user satisfaction survey should be used to find out how a given DW satisfies its user's expectations.

**Online Analytical Processing (OLAP)**

OLAP performs multidimensional analysis of data and provides the capability for complex calculations, trend analysis, and sophisticated data modelling with a short execution time. That provides the insight and understanding of data in multiple dimensions to enable end-users to make better decisions. The most popular OLAP operations on multidimensional data are roll-up (consolidation), drill down, slice and dice, and pivot (rotation). "Roll-up" performs data aggregation computed in many dimensions. "Drill down" is the inverse of roll-up operation allowing users to navigate from less detailed data to more detailed data. Slice and dice operations perform a selection on one or more dimension of the given cube, thus resulting in a sub-cube. Pivot rotates the data axes in view in order to provide an alternative presentation of the data.

The OLAP systems are categorised into three types: (1) Relational OLAP (ROLAP) which uses relational or extended-relational database and does not require pre-computation; (2) Multidimensional OLAP (MOLAP) which uses array-based multidimensional storage engines for multidimensional views of data and often require pre-processing to create data cubes; (3) Hybrid OLAP (HOLAP) which is a combination of both ROLAP and MOLAP. It offers the higher scalability of ROLAP and the faster computation of MOLAP.

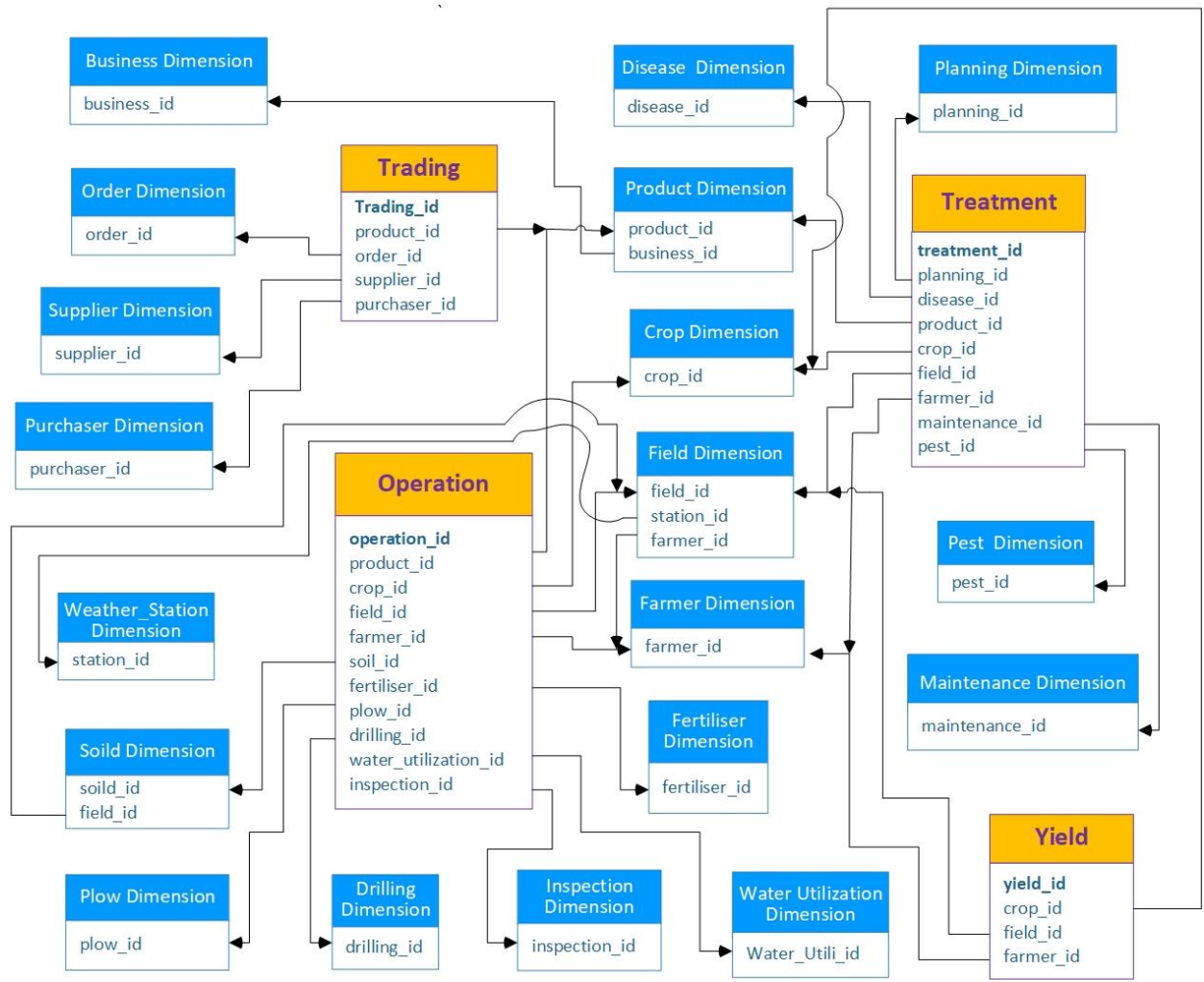

Figure 2. A part of our data warehouse schema for Precision Agriculture

In our Big Data context, ROLAP is not suitable because of its performance; each ROLAP report is an SQL query in the relational database that requires a significant execution time. Besides, ROLAP with its SQL statements does not meet all the users' needs, especially when performing complex calculations. While, MOLAP is a more traditional form of OLAP engine for data warehouses because it overcomes the disadvantages of ROLAP and the data warehouse is often built on a multi-dimensional schema. However, MOLAP has a disadvantage in that all calculations need to be performed during the data cube construction. Apache Kylin[1] is a distributed analytics engine that provides an extremely fast MOLAP engine at scale with more than 10 billion rows of data records. This open source engine implements some methods to ameliorate the above disadvantage. These methods are: (1) storing pre-calculated results to serve analysis queries; (2) generating each level's cuboids with all possible combinations of dimensions; (3) calculating all metrics at different levels; and (4) leveraging distributed computing power. However, the current MOLAP technology of Kylin is not really efficient in querying a high cardinality dimension. In addition, the combination of real-time/near-real-time and historical results for decision-making is really necessary but MOLAP is only good to serve queries on historical data. Hence, with agricultural Big Data, HOLAP should be able to exploit in-memory technologies of ROLAP to process data in real time.

### 4.4. High Dimensionality and Schema

The DW uses schema to logically describe the entire datasets. The DW schema consists of some fact tables and their corresponding dimension tables and their dependencies. A dimension is a structure that categorises data attribute, such as facts and observations (Oracle 9i.2 - 2007). The dimensions have some key functions to provide filtering, grouping and labelling. The rows in every dimension table are uniquely identified by a single key field. A fact table consists of dimension keys and measurements or metrics of a particular business process.

In the above schema of the collected datasets, the constellation schema, also known galaxy schema, contains many fact tables and dimension tables. We only present some particular fact and dimension tables in this paper. Figure 2 describes a part of constellation schema for PA which includes four fact tables and 19-dimension tables. The Fact tables are Trading, Operation, Treatment and Yield. The Trading fact table has four dimensions, namely Product, Order, Supplier and Purchaser. The Operation fact table has 10 dimensions, namely Product, Crop, Field, Farmer, Soil, Fertiliser, Plow, Drilling, Water_Utilization and Inspection. The Treatment fact table has 8 dimensions, namely Planning, Disease, Product, Crop, Field, Framer, Maintenance and Pest. Finally, the Yield fact table has three dimensions: Crop, Field and Farmer. Every fact table has primary key being combination of primary key of its dimension tables.

Table 1 describes some particular attributes of 19-dimension tables which are presented in Figure 2 and some dimensions of the DW. Each dimension table has a primary key with the same name of the table. There are some dimension tables shared between fact tables. For example, Crop, Farmer and Field dimension tables are shared by Operation, Treatment and Yield fact tables. Some dimension tables are linked together, such as Field dimension and Soil dimension.

---

[1] http://kylin.apache.org/

  

Table 1. Descriptions of dimension tables

| No. | Dimension tables | Particular attributes | Linking to Facts/Dimensions |
|---|---|---|---|
| 1 | Business | business_id, business name, original name, address, contact phone, contact mobile, contact email | Product dim. |
| 2 | Crop | crop_id, name, code, variety name, variety description, standard moisture percentage, estimated yield | Operation fact, Treatment fact, Yield fact |
| 3 | Disease | disease_id, name, type, features on crop, description, measure | Treatment fact |
| 4 | Drilling | drilling_id, method, machine, description, date | Operation fact |
| 5 | Farmer | farmer_id, name, sex, birth year, address, field area, phone, email, experiences, skills | Operation fact, Treatment fact, Yield fact, Field dim. |
| 6 | Field | field_id, station_id, farmer_id, name, block, area, working area | Operation fact, Treatment fact, Yield fact, Farmer dim., Weather_Staion dim. |
| 7 | Fertiliser | fertilizer_id, fertiliser nutrient, product name, application area, quantity, stock date | Operation fact |
| 8 | Inspection | inspection_id, description, problem type, severity, date, growth stage | Operation fact |
| 9 | Maintenance | maintenance_id, rate, description, date | Treatment fact |
| 10 | Order | order_id, order date, transaction date, value, comment, reference | Trading fact |
| 11 | Pest | pest_id, common name, scientific name, type, description, density, coverage | Treatment fact |
| 12 | Planning | planning_id, name, plan number, product name, product rate, date, water volume, notes | Treatment fact |
| 13 | Plow | plow_id, tillage method, plowing depth, machine, date | Operation fact |
| 14 | Product | product_id, product name, group name, type name, date of manufacture, business_id | Trading fact, Operation fact, Treatment fact, Business dim. |
| 15 | Purchaser | purchaser_id, name, address, contact person, contact phone, contact mobile, contact email | Trading fact |
| 16 | Soil | soild_id, field_id, mineral particles, organic matter, colour, PH value | Operation fact, Field dim. |
| 17 | Supplier | supplier_id, name, address, contact person, contact phone, contact mobile, contact email | Trading fact |
| 18 | Water_Utilization | water_utili_id, amount, source, method, date | Operation fact |
| 19 | Weather_Station | station_id, station name, station batch, measure date, air temperature, soil temperature | Field dim. |

## 5. Conclusion

In this paper, we described a data warehouse solution for crop farming. The data warehouse (DW) supports Big Data analytics and storage. The recent applications of precision agriculture systems and major challenges into building a successful precision agriculture data warehouse were reviewed and analysed. The DW architecture includes necessary modules to deal with large scale and efficient analytics. The presented schema herein was optimised for the datasets that were made available to us. Moreover, it is flexible and adaptable to other datasets. It has been designed in a constellation schema in order to facilitate quality criteria. Finally, descriptions and relationships of particular fact and dimension tables ware also modelled and included in the schema.

In future works, we will study and compare features of popular data warehouse software tools, such as Amazon Redshift, Google Mesa, Facebook Hive, MongoDB and Cassandra. We will select suitable technologies to implement the proposed data warehouse. The technologies selected will need to have some key characteristics, such as high performance, support for data science, high storage capacity, scalability and security. Next, we will conduct experiments on scalability of the proposed data warehouse on real-world datasets. We will also look at how to integrate this data warehouse with the high-performance knowledge map framework (Le-Khac *et al.*, 2007).

**Acknowledgements.**

This research is part of the CONSUS research programme which is funded under the SFI Strategic Partnerships Programme (16/SPP/3296) and is co-funded by Origin Enterprises Plc.